\documentstyle[12pt,epsfig]{article}
 1
\def\as{\alpha_{\rm S}}

\def\citenum#1{{\def\@cite##1##2{##1}\cite{#1}}}
\def\citea#1{\@cite{#1}{}}
\def\as{\alpha_{\rm S}}

\def\g{\gamma}

\def\l{\lambda}

\def\s{\sigma}

\def\({\left(}
\def\){\right)}

\def\citenum#1{{\def\@cite##1##2{##1}\cite{#1}}}
\def\citea#1{\@cite{#1}{}}

\def\l1vt{\vec{l_{1\perp}}}

\def\bt{b_{\perp}}

\def\bt2{$b^2_t$}

\def\jol1{$J_0(\,l_{1\perp}\,r_{\perp}\,)$}

\def\citea#1{\@cite{#1}{}}

\def\beq{\begin{equation}}
\def\eeq{\end{equation}}
\def\bea{\begin{eqnarray}}
\def\eea{\end{eqnarray}}

\def\eq#1{{Eq.~(\ref{#1})}}

\def\bbbz{{\mathchoice {\hbox{$\sf\textstyle Z\kern-0.4em Z$}}
{\hbox{$\sf\textstyle Z\kern-0.4em Z$}}
{\hbox{$\sf\scriptstyle Z\kern-0.3em Z$}}
{\hbox{$\sf\scriptscriptstyle Z\kern-0.2em Z$}}}}
\def\npb#1#2#3{    {\it Nucl. Phys. }{\bf B#1} (19#2) #3}
\def\plb#1#2#3{    {\it Phys. Lett. }{\bf B#1} (19#2) #3}
\def\prd#1#2#3{    {\it Phys. Rev. }{\bf D#1} (19#2) #3}

\def\zpc#1#2#3{    {\it Z. Phys. }{\bf C#1} (19#2) #3}

\def\sjnp#1#2#3{   {\it Sov. J. Nucl. Phys. }{\bf #1} (19#2) #3}

\relax
\begin{document}
\begin{titlepage}
\noindent
\begin{flushright}
January  1997 \\ 
TAUP 2405/97
\end{flushright}
\vspace{1cm}
\begin{center}
{\Large\bf BEHAVIOUR OF THE FORWARD PEAK}\\[2ex]
{\Large \bf  IN HARD DIFFRACTIVE LEPTOPRODUCTION}\\[2ex]
{\Large \bf OF VECTOR MESONS}\\[4ex]
{\large E. G O T S M A N${}^{a), 1)}$, E. L E V I N${}^{a),b),2)}$\,\,
 and U.\,\,M A O R${}^{a), 3)}$}
 \footnotetext{$^{1)}$ Email: gotsman@post.tau.ac.il .}
\footnotetext{$^{2)}$ Email: leving@ccsg.tau.ac.il.}
\footnotetext{$^{3)}$ Email: maor@ccsg.tau.ac.il.}\\[4.5ex]
{\it a) School of Physics and Astronomy}\\
{\it Raymond and Beverly Sackler Faculty of Exact Science}\\
{\it Tel Aviv University, Tel Aviv, 69978, ISRAEL}\\[1.5ex]
{\it b)  Theory Department, Petersburg Nuclear Physics Institute}\\
{\it 188350, Gatchina, St. Petersburg, RUSSIA}\\[3.5ex]
\end{center}
~\,\,\,
\vspace{2cm}

{\large \bf Abstract:}

The measured forward slope in elastic and inelastic leptoproduction
of  vector mesons differ by a substantial amount.
In an attempt to describe this phenomenon we construct a two radii model
for the target proton, and estimate 
 the effective
parameters of the hard Pomeron obtained from a pQCD dipole model with
eikonal shadowing corrections (SC). We show that the SC reduce the
intercept of the hard
Pomeron and generate an effective shrinkage of the forward peak and a
diffractive dip at $|t| \approx 1 GeV^2$, which appears to affect the
value of the experimentally measured slope.

\end{titlepage}

\section{Introduction}

\indent Perturbative QCD (pQCD) calculations of diffractive leptoproduction
of vector mesons (DLVM) in deep inelastic scattering (DIS) are based on
the observation \cite{RYS,KOPEL,BROD} that  DLVM is a hard process, 
dominated by short 
transverse distances $r_{\perp}^{2} \propto \frac{1}{Q^{2}}$. 
In contrast, elastic photoproduction of light vector mesons has the typical
properties of soft interactions,  i.e. a pronounced forward peak shrinking
with
energy, and a mild energy dependence of the integrated cross section. The  
forward peak of  DLVM in DIS is less distinct while the energy 
dependence of the integrated cross section is rather 
steep\cite{ZEUSVM,H1VM}. The same 
properties are seen in real photoproduction of  J/$\psi$
\cite{ZEUSVM,H1VM}
where $ 4 
m^2_{\psi}$ replaces $Q^2$ as a measure of the hard scale \cite{RYS}.
These experimental observations justify a pQCD calculation
\cite{GLMVM,FKS,RRML} in 
which $x$ dependence of calculated cross section is determined by
 $[xG(x,Q^{2})]^{2}$, where 
$xG(x,Q^{2})$ denotes the gluon distribution within the target.

The goal of this letter is to examine the detailed behaviour of the 
forward DLVM cross section.
 To this end we shall utilize  the expotential
approximation for
\beq \label{EXP}
\frac{d \s}{d t}\,\,=\,\,\(\frac{d \s}{d t}\)_{t=0}\,e^{-\,B\,|t|}\,\,,
\eeq
 where the $t$ = 0  slope  is defined by
\beq \label{B}
B = \frac{\frac{d}{dt}(\frac{d \sigma}{dt})}{\frac{d\sigma}{dt}}|_{t=0}
\eeq
\indent The properties of the differential cross section in 
 high energy soft hadronic processes are well understood,  these are
dominated by Pomeron
exchange, where we have
\beq
\alpha(t) = \alpha_P(0) + \alpha^{'}_Pt = 1 + \epsilon + \alpha^{'}_P t\,\,.
\eeq
The energy dependence of the forward differential cross section is 
determinated by $\epsilon$, whereas the slope $ B = B_{0} + 2 \alpha^{'}_P 
ln( \frac{s}{s_{0}}) $ depends on $\alpha'_P$.
 A good reproduction of
  the hadronic data is obtained  using  the 
Regge parameterization \cite{DL,BKW},  with $\epsilon$ = 0.08 and
$\alpha^{'}_P$ = 0.25 GeV$^{-2}$. The observation that $\alpha^{'}_P \neq
$ 0,
is associated with the experimentally observed shrinkage of the forward
differential elastic cross section, with increasing energy ( $s = W^2$ ).
It is very convenient to parameterize the hard Pomeron in a similar fashion,
where $\epsilon (Q^2)$ and $\alpha'_P(Q^2)$ provide the relevant 
information on the $W$ and $t$ dependence of the DLVM cross section.

\indent An important property, observed in $pp$ and 
${\bar p}p$ 
elastic scattering in the ISR-Tevatron range, is a diffractive dip of
$\frac{d\sigma}{dt}$  positioned at $|t| \approx$ 1.4 GeV$^{2}$
at the lower ISR energies, which moves slowly towards smaller  $t$  as the 
energy is increased.
 This phenomenon has been associated with screening
re-scattering corrections, which are calculated with relative ease in the 
eikonal approximation \cite{CY,BSW,GLM}. An obvious consequence of a
dip at higher $|t| $ values, is that,  the exponential approximation for
$\frac{d\sigma}{dt}$  is only  valid  in the narrow forward cone, as
B acquires a positive curvature outside this cone.

In the following we study the dependence of the forward DLVM cross 
section on $W, t$ and $Q^2$. Specifically we wish to examine:\\
1) The $Q^2$ dependence of $\epsilon$ in hard DLVM.\\
2) The phenomena of shrinkage and a diffractive dip in the wide forward 
cone of hard DLVM, as  a function of $W$ and $Q^2$.

We note that for the hard Pomeron in  the DGLAP approach    $\epsilon$ grows 
with
$Q^2$, and there is no shrinkage or diffractive dips. We re-examine these 
features following our paper \cite{GLMVM} and utilize a screened dipole model
in the DGLAP evolution equations, which have been shown to be adequate for 
the HERA kinematical range. Furthermore, we note that the measured slopes
in elastic and inelastic DLVM differ by a substantial amount \cite{HERAPSI}.
This suggests the possibility that there are two intrinsic radii in the 
target proton. To investigate this possibility, we have constructed an 
appropriate model with two radii which takes into account the SC. Our 
calculated results are compared with the available experimental data. 
We are successful in reproducing  the experimental results for J/$\psi$,
however, for $\rho$ production our pQCD
calculations  do not agree with data. We attribute this deficiency to the 
presence of a non negligible soft Pomeron component, and discuss it briefly.

\section{Main formulae}
A general approach to the SC for DLVM in DIS, has been developed in
Ref.\cite{GLMVM}, following
 Refs.\cite{LERY}\cite{MU90}. It was shown \cite{GLMVM} that  the
  diffractive production of vector mesons occurs at small distances,
  and hence the passage of the quark -
antiquark pair through the target, can be calculated in pQCD utilizing the
eikonal approximation. 
 The  situation with the SC
in the gluon sector  turns out to be more complicated. However, in the
eikonal
approach the SC  can be absorbed 
 into gluon density.
The general formula   for the screened DLVM amplitude 
 \cite{GLMVM} is:
\beq \label{10}
A(b_t;Q,x)\,=\,C\,\,\int \,\frac{d^2 r_{\perp}}{ \pi}\,\int^1_0 \,dz
\Psi^{\g^*}(
Q,r_{\perp},z)\,2\,\{\,1\,-\,e^{-\,\kappa(r_{\perp},x;b_t)}\,\}\,
[\Psi^V(r_{\perp}= 0 ,x)]^*,
\eeq
where $\Psi^{\g^*}$,$\Psi^V$ denote the wave functions of the virtual
photon and vector meson respectively. The constant $C$ has been calculated
in Refs.\cite{RYS}\cite{BROD}\cite{GLMVM}. $r_{\perp}$ denotes the
transverse distance between  the quark and antiquark, and  $z$  the
fraction of
the photon energy carried by the  quark (antiquark). $W$ denotes the energy
and 
 $b_t$  the impact
parameter of the reaction. 
$x\,\,=\,\,\frac{Q^2\,+\,m^2_V}{W^2}$, where $m_V$ is the vector meson
mass. The degree of the SC in \eq{10} is characterized by the  parameter
$\kappa$ defined below.  In 
our model $\kappa$ depends on \\ the structure of the
target. 
\subsection{One radius approach.}

In the simplest  one radius model for the target structure, we have
\cite{GLMVM}:
\beq \label{11}
\kappa(b_t;r_{\perp},x)\,\,=\,\,\frac{2 \pi \as
r^2_{\perp}}{3}\,\Gamma(b_t)\,\,xG(x,\frac{4}{r^2_{\perp}})\,\,.
\eeq
Here, we have used the main property of the DGLAP evolution equations,
which allows us to factor out the $b_t$ dependence from  $x$ and
$r_{\perp}$  (see Ref.\cite{LR90}). $\Gamma(b_t)$ denotes the Fourier
transform of the two gluon form factor of the target, which has the same
form for any process.

The relation between the  profile $\Gamma(b_t)$ and the two gluon form
factor
reads
\beq \label{b1}
\Gamma(b_t)\,\,=\,\,\frac{1}{\pi}\,\,\int\,e^{- i (\vec{b}_t
\cdot\vec{q}_t)}\,F(t)\,d^2 q_t
\eeq
with $t\,=\,q^2_t$. To simplify the  calculations we approximate 
\beq \label{b2}
\Gamma(b_t)\,\,=\,\,\frac{1}{R^2}\,e^{- \frac{b^2_t}{R^2}}\,\,.
\eeq
Intergrating over $r_{\perp}$ in \eq{10} (see Ref.\cite{GLMVM}) we have
\beq \label{12}
A(b_t,Q,x)\,\,=\,\,C\,\int^1_0 d z
\,2\,\{\,1\,\,-\,\,Y(\kappa(x,b_t))\,\}\,\Psi^V(0,z)\,\,,
\eeq
where
\beq \label{Y}
Y(\kappa)\,\,=\,\,
\frac{1}{\kappa}\,e^{\frac{1}{\kappa}}\,E_1(
 \frac{1}{\kappa})
\eeq
with
\beq \label{KAPPA}
 \kappa(x,b_t)\,\,=\,\,\frac{2 \pi \as}{3\,a^2\,R^2
}\,e^{-
\frac{b^2_t}{R^2}}\,\,x G(x,a^2)\,\,,
\eeq
  $a^2\,=\,Q^2 z (1 - z)
\,+\,m^2_Q$, and $m_Q$ denotes the quark mass.

In general, the slope of the differential cross section at t = 0, is given
by:
\beq \label{13}
B\,\,=\,\,\frac{1}{2}\,\frac{\int \,A(b_t,Q,x)\,b^2_t\,d^2 b_t}{\int
\,A(b_t,Q,x)\,d^2b_t}
\eeq
while the t-dependence of the differential cross section is obtained from
\beq \label{14}
\frac{\frac{d \s}{d t}}{\frac{d \s}{d t}|_{t =
0}}\,\,=\,\,\left|\frac{\int J_0 (b_t\,\sqrt{|t|}) \,A(b_t,Q,x)\,d^2 b_t}{
\int \,A(b_t,Q,x)\,d^2 b_t}\right|^2\,\,.
\eeq
~\\
For the slope at $t = 0$, we can derive a simple formula based on
\eq{12},
namely
\beq \label{15}
B\,\,=\,\,\frac{R^2}{2}\,\,\frac{\int^1_0 \,d z \int^{\infty}_0
  \,du\,\frac{e^{- u}}{u}\,\ln(1 +
  u\,\kappa)\,\Psi_V(z,0)}{\int^1_0\,d z
  E_1(\frac{1}{\kappa})\,e^{\frac{1}{\kappa}}\,\Psi_V(z,0)}\,\,.
\eeq
In Figs.\ref{jslope} and \ref{rhoslope} we plot the values that we obtain
for the slope $B$, for different values of $Q^2$, as a function of the
energy, where we have taken $R^2\,=\,6\,GeV^{-2}$. For J/$\psi$
production we have $z$= 1/2 ( see Ref.\cite{GLMVM} for details),
while for $\rho$    production we need to integrate over $z$ using the
asymptotic wave function of the form $\Psi_{\rho}(z,0)\,=\,z(1 -
z)\,\phi_{\rho}(0)$.

\subsection{Two radii approach}

Recent data from HERA on J/$\psi$ production\cite{HERAPSI} has
some novel features, one is the fact that the elastic and inelastic photo
and lepto produced J/$\psi$, have completely different slopes ($t$ -
dependence) with $B_{el}$\,=\,4 \,$GeV^{-2}$ and $B_{in}$\,=\,1.66
$GeV^{-2}$ (see Fig. \ref{fig1} ), although the values of the integrated 
cross
section are the same. This suggests that  two different radii maybe
present  in the proton target. Consequently, we extend our formalism to a  two
radii model, with the aid of a generating function
\beq \label{16}
\s(\eta,x,r^2_{\perp};b_t)\,\,=\,\,\sum^{\infty}_{n = 0}\,( - 1)^n\,e^{-n\,
  \eta}\,\,C_n(x,r^2_{\perp};b_t)\,\,
\eeq
where $C_n$ denotes the amplitude for $n$ - Pomeron exchange. For
vector meson production the amplitude is:
\beq \label{17}
A(x,Q^2;b_t)\,\,=\,\,C\,\,\int \,\frac{d^2 r_{\perp}}{\pi}\,\int^1_0
\,d z
\Psi_{\g^*}(Q,z,r_{\perp})\,\,\s(\eta=0,x,r_{\perp};b_t)\,\Psi_V(r_{\perp}=0,z)\,\,.
\eeq
A trivial calculation shows that \eq{10} is a solution of the
equation
$$
e^{- \eta}\,\frac{d \s}{d \eta}\,\,=\,\,\kappa\,(\,1\,\,-\,\,\s\,)\,\,,
$$
as  illustrated in Fig. \ref{fig5}a. For the two radii model we
obtain a  set of equations which are illustrated in Figs.\ref{fig5}b-d.
$$
e^{- \eta}\,\frac{d \s^e_e}{d
  \eta}\,\,=\,\,\kappa_1\,(\,1\,-\,\s\,)\,\,-\,\,\kappa_2\,\s^i_e\,\,;
$$
$$
e^{- \eta}\,\frac{d \s^i_e}{d
  \eta}\,\,=\,\,\kappa_2\,(\,1\,-\,\s^i_i\,)\,\,-
\kappa_1\,\s^i_e\,\,;
$$
\beq \label{18}
e^{- \eta}\,\frac{d \s^i_i}{d
  \eta}\,\,=\,\,\kappa_2\,(\,1\,\,-\,\,\s^i_i\,)\,\,.\\ 
\eeq
In \eq{18} we have assumed the same values of $\kappa$ for a transition of a
proton to inelastic state, and for an inelastic state to an inelastic
state. This is true in the additive quark model and we consider  it
a reasonable first approximation here. $\kappa_1$ and $\kappa_2$ are
defined as follows:
\beq \label{kappas}
 \kappa_1(x,b_t)\,\,=\,\,
 \frac{2 \pi \as}{3\,a^2\,R^2_1}
e^{-\,\frac{b^2_t}{R^2_1}}\,\,x G(x,a^2)\,\,;
\eeq
$$
 \kappa_2(x,b_t)\,\,=\,\,\frac{2 \pi \as}{3\,a^2\,R^2_2}\,v\,
\,e^{-
\frac{b^2_t}{R^2_2}}\,
\,x G(x,a^2)\,\,;
$$
with $v \,=\,\frac{R_1}{R_2}$ whose value  is taken from the experimental
data,
which shows the same cross section for elastic and inelastic diffractive
production of J/$\psi$ meson \cite{HERAPSI}.

The solution to \eq{18} is
$$
\s^e_e(\eta=0,x,r_{\perp};b_t)\,\,=\,\,1 \,\,-\,\,e^{-
  \kappa_1}\,\,+\,\,
\frac{\kappa^2_2}{\kappa_1\,-\,\kappa_2}\,e^{-
  \kappa_1}\,\,+
$$
\beq \label{SOL}
\,\,\frac{\kappa^2_2}{(\kappa_1\,-\,\kappa_2)^2}\,(\,e^{- \kappa_1}\,
\,-\,\,e^{- \kappa_2}\,)\,\,;
\eeq
and
\beq \label{SOL1}
\s^i_e(\eta
=0,x,r_{\perp};b_t)\,\,=\,\,\frac{\kappa_2}{\kappa_1\,-\,\kappa_2}
\,\{\,e^{- \kappa_2}\,\,-\,\,e^{- \kappa_1}\,\}\,\,.
\eeq
~\\
Using the technique developed in Ref.\cite{GLMVM} we can integrate
over $r_{\perp}$ and obtain:
$$
A^e_e(x,Q^2;b_t)\,\,=\,\,C\,\,\int^1_0\,d z \,\,\{\,\,1\,\,-\,\,Y(\kappa_1)
\,\,-\,\,\frac{\kappa^2_2}{\kappa_1(\kappa_1\,-\,\kappa_2)}\,
[\,\frac{1}{\kappa_1}\,(\,1\,-\,Y(\kappa_1)\,)\,-\,Y(\kappa_1)\,]\,\,-
$$
\beq \label{21}
\,\,\frac{\kappa^2_2}{(
\kappa_1\,-\,\kappa_2)^{2}}\,[\,Y(\kappa_2)\,-\,Y(\kappa_1)\,]\,\}
\frac{\Psi_V(0,z)}{a^2}\,\,,
\eeq
and
\beq \label{22}
A^i_e(x,Q^2;b_t)\,\,=\,\,C\,\,\int^1_0\,dz\,\frac{\kappa_2}{\kappa_1
  \,-\,\kappa_2}\,\,\{\,Y(\kappa_1)\,\,-\,\,Y(\kappa_2)\,\}\,
\frac{\Psi_V(z,0)}{a^2}\,\,.
\eeq

 \section{Results and Discussion}

\indent The most important parameters in our formalism are  $\kappa_{i}$
 (with i =1,2, see
\eq{21}),  they  depend on the form assumed for the profile function
$\Gamma(b_{t})$, and on the value of the gluon density
$xG(x,Q^{2})$. Our formalism for the  SC only includes screening  due to
the propagation 
of the ${\bar q}q$ pair through the target. We assume that the SC in the gluon
 sector, have been  incorporated in the parametrization used for
the gluon density, which has been determined by fitting to
the 
experimental data. In all our calculations we have used the GRV
parametrization \cite{GRV}, which successfully describes the experimental
data on
$F_{2}(x,Q^{2})$ down to small photon virtualities, $Q^{2} \approx$ 1.5
 GeV$^{2}$.
\par
\begin{figure}
\epsfig{file=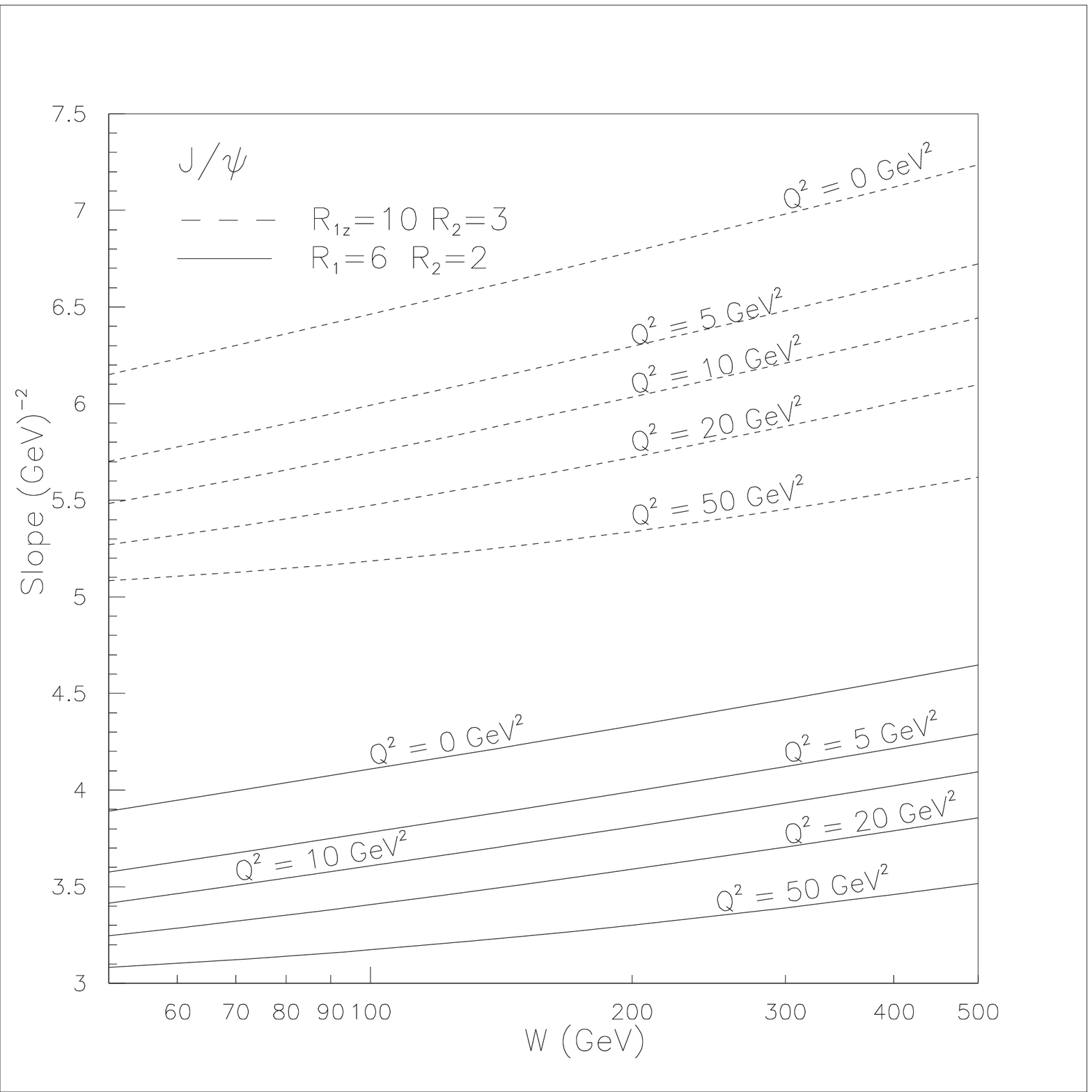,width=160mm}
\caption{$W$ and $Q^2$ behaviour of the slope at $t = 0$ for J/$\psi$
diffractive production, for the two sets of radii.} 
\label{jslope}
\end{figure}

 Our parametrization depends on the values taken for the
radii $R_{1}^{2}$ and $R_{2}^{2}$, which describe the t ($b_{t}$)
dependence of the two gluon form factor at the elastic and inelastic
proton vertex. We have chosen $R_{1}^{2}$ = 6 GeV$^{-2}$ and
$R_{2}^{2}$ = 2 GeV$^{2}$, these values reproduce the experimental data  for
the t-slope in elastic and inelastic photoproduction of $J/\psi$
($ \gamma p \rightarrow J/\psi p)$ and $ ( \gamma p \rightarrow J/\psi X)$ 
i.e. $B_{el}$ = 4.0$ \pm$ 0.3  GeV$^{-2}$ and  $B_{in}$ = 1.6 $\pm$ 0.3
 GeV$^{-2}$
at W = 100 GeV (see \cite {H1DATA}).
 For $J/\psi$ production, typical transverse distances are of
the order $r_{\perp}^{2} \approx (\frac{Q^{2}}{4} + m_{c}^{2})^{-1}$, 
which justifies
the use of pQCD. 
 For this reason we used the $J/\psi$ diffractive data to fix our 
parameters. Our
results for $J/\psi$ are plotted in Fig.\ref{jslope},   which
displays 
the W and
$Q^{2}$ dependence of the effective slope. 
 As expected, SC change the dependence of $B$ on the kinematical variable.
For 
example, for
W = 100 GeV and $Q^{2}$ = 0, the slope is B = 4 GeV$^{-2}$, while 
at large
$Q^{2}$ ($Q^{2}$ = 50 GeV$^{2}$) and the same energy, the slope decreases
to a
 value B = 3.1 GeV$^{-2}$.
 The energy dependence of the slope can be parametrized by an effective
$\alpha^{'}_{eff}$, using \eq{B} we can write the slope in the form
\beq
B = B_{0} + 4 \alpha_{eff}^{'} ln(\frac{W}{W_{0}})
\eeq
For $Q^{2}$ = 0, in the energy range 50 $\leq$  W $\leq$ 500 GeV, we
obtain a value 
$\alpha_{eff}^{'}$ = 0.08 GeV$^{-2}$. For large values of $Q^{2}$,
 $\alpha_{eff}^{'}$
decreases. Our value for the slope at $Q^{2}$ = 0, is approximately
 one fourth of the conventional
value assumed for the slope of the soft Pomeron.
\par
 We wish to stress, 
that both experimental and theoretical values
quoted for the slope are  strongly dependent on the manner in which
the  t-slope, is determined.
To investigate this ambiguity, we have  calculated the t dependence 
of the
 differential
cross section 
for $J/\psi$ production for different intervals of $t$. In Fig.\ref{jtslope}
 we  clearly see 
 the effect of the SC in changing the exponential dependence in t.
 It gives rise to   a positive curvature at small t, and a dip at
$|t| \approx$ 1 GeV$^{2}$. The position of the dip is energy 
dependent, and at
high energies the dip moves to smaller values of $|t|$. One
should  therefore  take 
heed when 
attempting to determine the slope from measurements made over a wide range of t.
For example, for $Q^{2}$ = 0 and W = 50 GeV we obtain a
 $B_{eff}$ = 4.6 GeV$^{-2}$, when determining the slope over the range
from t = 0 to $|t|$  = 0.8 GeV$^{-2}$. This is to be compared with
$B_{eff}$
=
3.9 GeV$^{-2}$ at t = 0 (see  Fig.\ref{jslope}).
 The energy dependence of the slope determined
from the 0 $\leq |t| \leq$ 0.8 $GeV^2$  distribution gives a
$\alpha_{eff}^{'}$ = 0.19 GeV$^{-2}$,
which is close to the value for the soft Pomeron.
  \par
\begin{figure} 
\epsfig{file=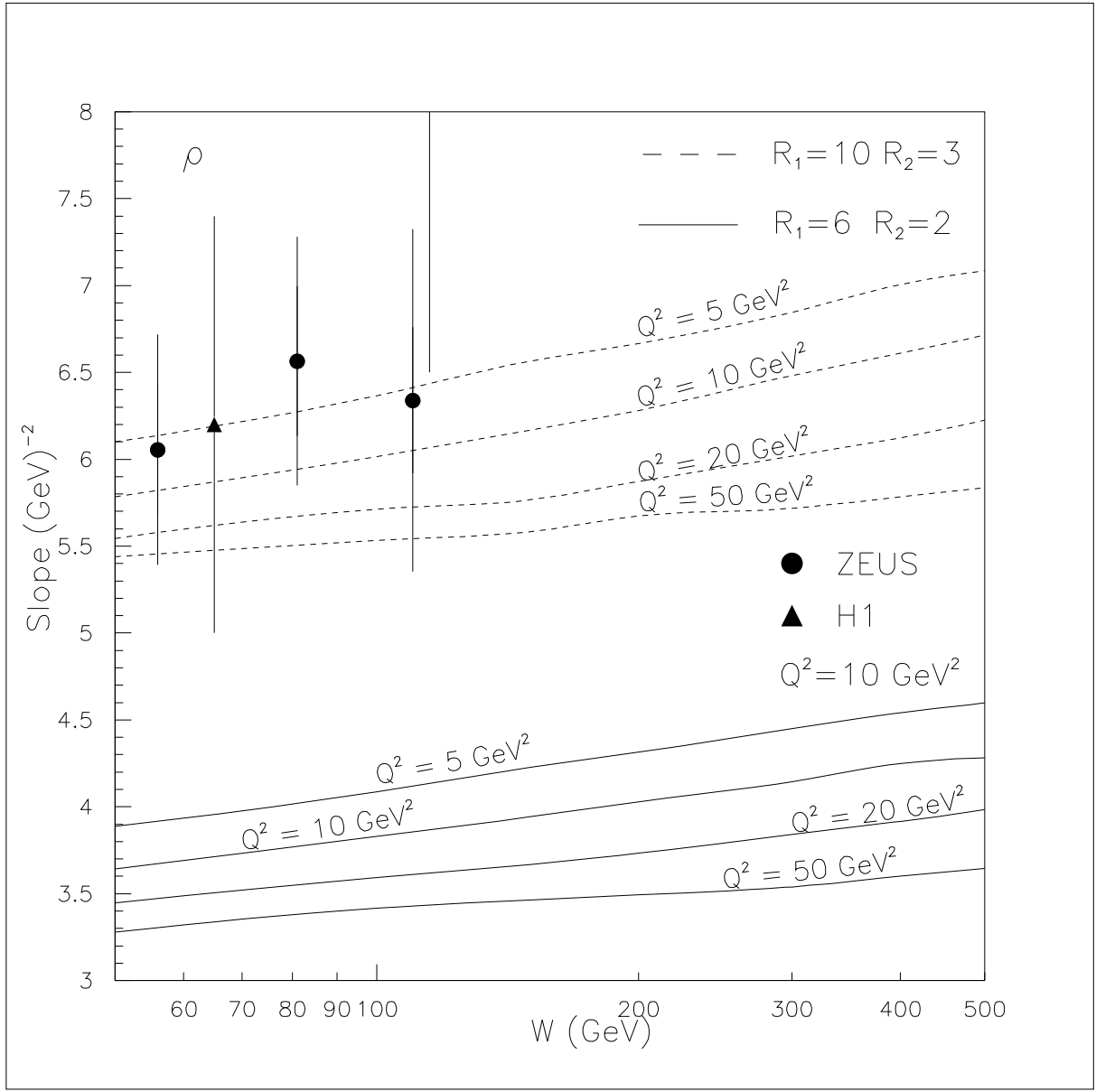,width=160mm}
\caption{$W$ and $Q^2$ dependence of the slope at $t=0$ for $\rho$ diffractive
production, for the two sets of radii.}
\label{rhoslope}
\end{figure}

\begin{figure}
\epsfig{file=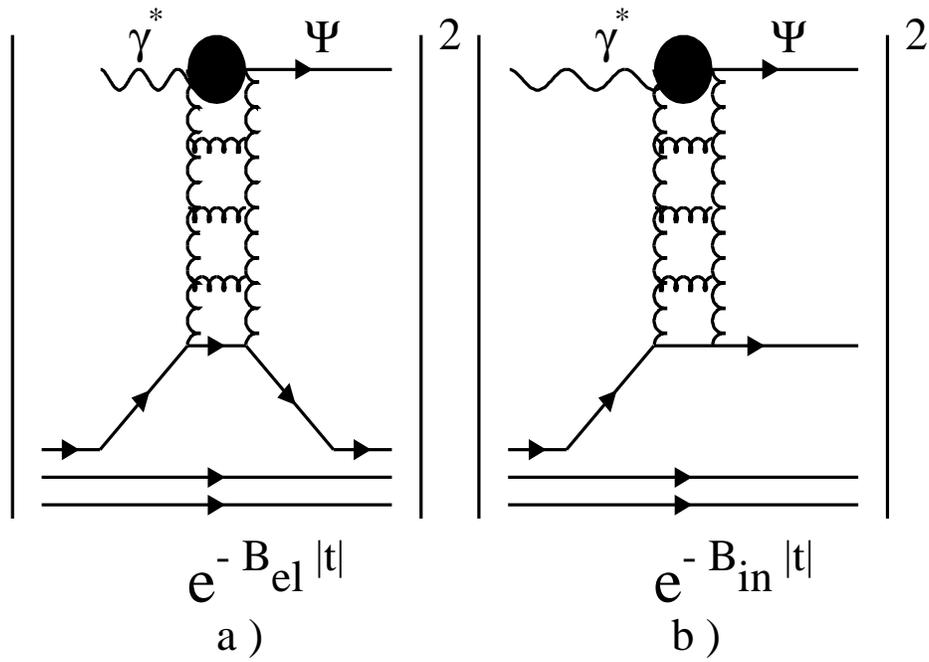,width=160mm}
\caption{The J/$\psi$ production without (a) and with ( b)
proton dissociation.}
\label{fig1}
\end{figure}

\begin{figure}
\epsfig{file=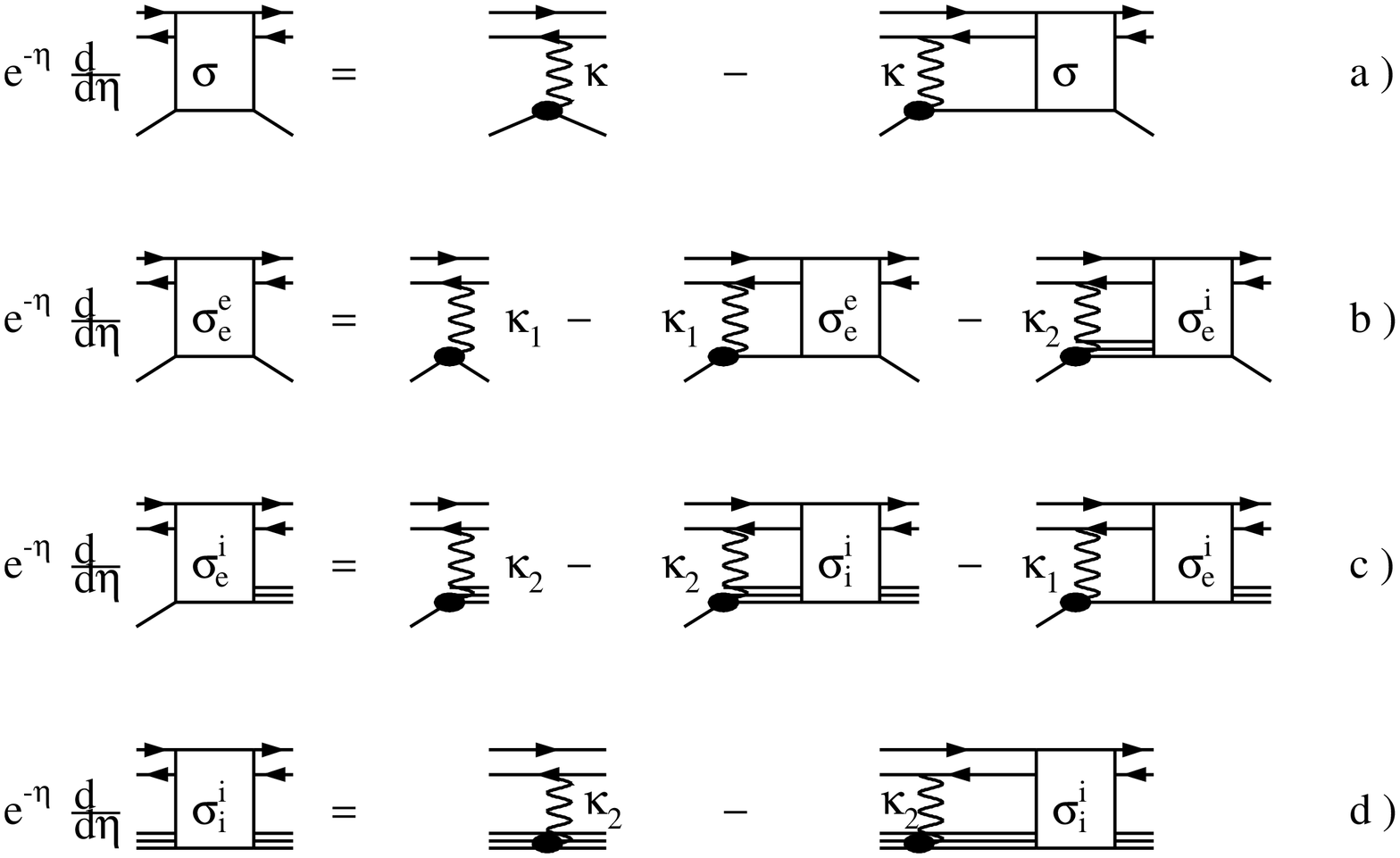,width=160mm}
\caption{Equations for the SC in eikonal approach in the two radii model
of the target proton.}
\label{fig5}
\end{figure}

\begin{figure}
\epsfig{file=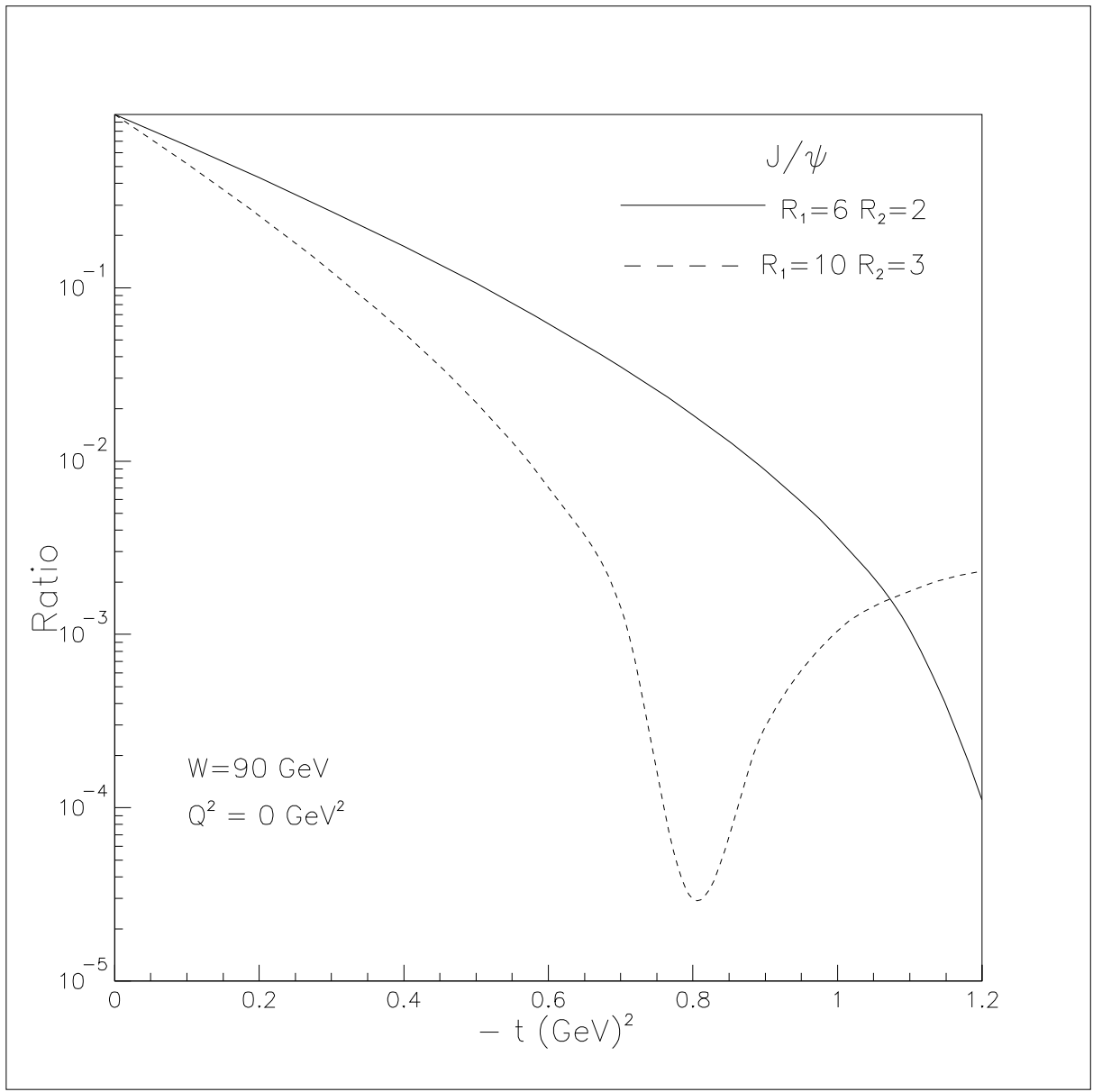,width=160mm}
\caption{$t$-dependence of the differential cross section for diffractive 
J/$\psi$ production.}
\label{jtslope}
\end{figure}

 The t slope,
the scale of the effective slope, as well as $\alpha_{eff}^{'}$ depend on
the values taken for the radii. This is illustrated in Figs.\ref{jtslope},
\ref{rhoslope} and \ref{rhoqslope}. These quantities also differ for the
one and two radii models. Taking $R_{1}^{2}$
 = 10 GeV$^{-2}$
and $R_{2}^{2}$ = 3 GeV$^{-2}$ produce a minimum at values of 
$|t| <$ 1 GeV$^{2}$. 
\par
For $\rho$ production the case is more complicated, as we are not sure
that we are dealing only with a hard process \cite{HERAPSI}.
Our results for  $\rho$ production are  qualitatively  the same as those
for J/$\psi$
 (see Fig. \ref{rhoslope}). However,
 for  $\rho$ production we fail to reproduce the experimentally measured
value of 
B = 6 GeV$^{-2}$ ( at $Q^{2} = 5-10 \; GeV^{2}$), as  we obtain a value\\
 B = 4 GeV$^{-2}$.
This is not suprizing, as our approach is based on pQCD, and in
 $\rho$ production for values of
$Q^{2}$ = 5 - 10  GeV$^{2}$, we still expect a non-perturbative component, 
 which we have not taken into account, to present. At larger values of
$Q^{2}$ (where one can rely exclusively  on pQCD) the experimental slope
\cite{HALINA}
approaches the value  B=3 - 4 GeV$^{-2}$. We find that $\alpha_{eff}^{'}$
is small in the region of large $Q^{2}$ ($\approx$ 0.05 GeV$^{-2}$
at t=0).
\par In \eq{B}, we have parametrized the profile function $ \Gamma(b_{t})$ 
by an exponential form. For small t where our result is rather insensitive
to the form taken for $\Gamma(b_{t})$, the exponential form is 
adequate. At larger $|t|$  this is no longer  so. We have checked 
the influence
 of taking a different form for $\Gamma(b_{t})$
on the t-distribution. 
For the case where 
 $$ \Gamma(b_{t}) = \frac{2}{R^{2}}(2\sqrt{2}\frac{b_{t}}{R})K_{1}(2\sqrt{2}
\frac{b_{t}}{R})$$
(where $K_{1}(x)$ denotes the associated Bessel function of first order),
this corresponds to assuming a dipole for the two gluon form factor
$$ F(t) = ( 1 + \frac{R^{2}|t|}{8})^{-2} $$
We find that there is no change in the energy dependence of the slope
at t=0. However, the dipole form factor does effect the t distribution for
$|t| \geq$ 0.3 GeV$^{2}$ , making it shallower than the one obtained with
an exponential form factor. Our result that 
$\alpha_{eff}^{'} \approx \alpha_{P}$ (soft Pomeron) over a wide range of
t, still holds.

We  notice that the power -like form factor, given in the 
previous equation, leads to a slope which  depends  strongly on the 
region of $t$ where the slope is measured. Indeed, using this equation we have
\beq
B(t=t_0)\,\,=\,\,\frac{R^2}{2( 1 + \frac{R^2}{8} \,|t_0|)}\,\,.
\eeq
With $|t_0| = 0.1 GeV^2$ and $R^2 = 10 GeV^{-2}$ we have  $B(|t|= 0.1)\,=\,
0.8
\,B(t=0)$.
Therefore, part of the $Q^2$ dependence of the experimental slope, could be
attributed to the different values of $t_{min}$, at which the slope was 
measured.

\indent 
\section{Conclusions}
~\\
1. Based on the new experimental data from HERA \cite{HERAPSI},
we have assumed 
a two radii model for the proton. We have obtained formulae for shadowing
 screening  corrections for diffractive vector meson production in DIS.
This has been calculated within the framework of the DGLAP evolution
equations for the low $x$ region.
~\\
2. The influence of the SC on the absolute value, the $Q^{2}$ and W
dependence of the t-slope for diffractive vector production in DIS,
is discussed. We show that SC induces a shrinkage of the t slope.
We also calculated the value of the effective $\alpha^{'}$.
~\\
3. We  summarize the results of our calculations in Table I, which shows
the values of the intercept and the slope of the effective trajectory of
the hard
Pomeron, that we obtain from our calculations. Comparing the trajectories
of the hard  Pomeron obtained without and with SC, we note that the 
SC reduces the value of the intercept and generates an effective 
$\alpha^{'}_{eff}$ for the hard Pomeron.
~\\
4. We show that the SC changes the t-distribution, creating a positive
curvature at small values of t, and a dip at $|t| \approx$ 1 GeV$^{2}$.
These two phenomena suggest,  that one should be  careful  when attempting
to determine the t-slope from data taken over
a wide range of t. The slope 
determined under such conditions,  suggest a shrinkage which is much 
larger than the slope at t=0, given in Table I.
~\\
5. We have chosen the values of the two radii to reproduce the
experimental
data for $J/\psi$ diffractive production, where one can safely
apply pQCD. However, these values fail to reproduce the value of the t-slope
for $\rho$ production in the region of $Q^{2}$ = 5 - 10 GeV$^{2}$.
The results of our phenomenological investigation support the conclusions
of the H1 collaboration \cite{HERAPSI}, in that hard  physics
 effects already dominate
 at very small $Q^{2}$ for J/$\psi$ production, while for $\rho$
production even at values of $Q^{2}$ = 10 - 15 GeV$^{2}$, 
  a non negligible soft component still appears to be
present. Direct evidence for this is the experimentally observed shrinkage
of the forward elastic peak for $\rho$ production, which is much larger than
 that  expected for  only  hard
processes \cite{FKS} (even when  SC are included).
~\\
6. Based on our investigation we recommend  
that: \\
(i) The diffractive t-slope be determined from measurements made for
$|t | \leq$ 0.3 GeV$^{2}$. \\
(ii) For $\rho$ production the hard diffractive slope should be
 measured  for $ Q^{2} \geq$ 15 GeV$^{2}$, where we
expect $B_{diff} \approx$ 4 (GeV)$^{-2}$, or smaller.

\newpage

\begin{figure}
\epsfig{file=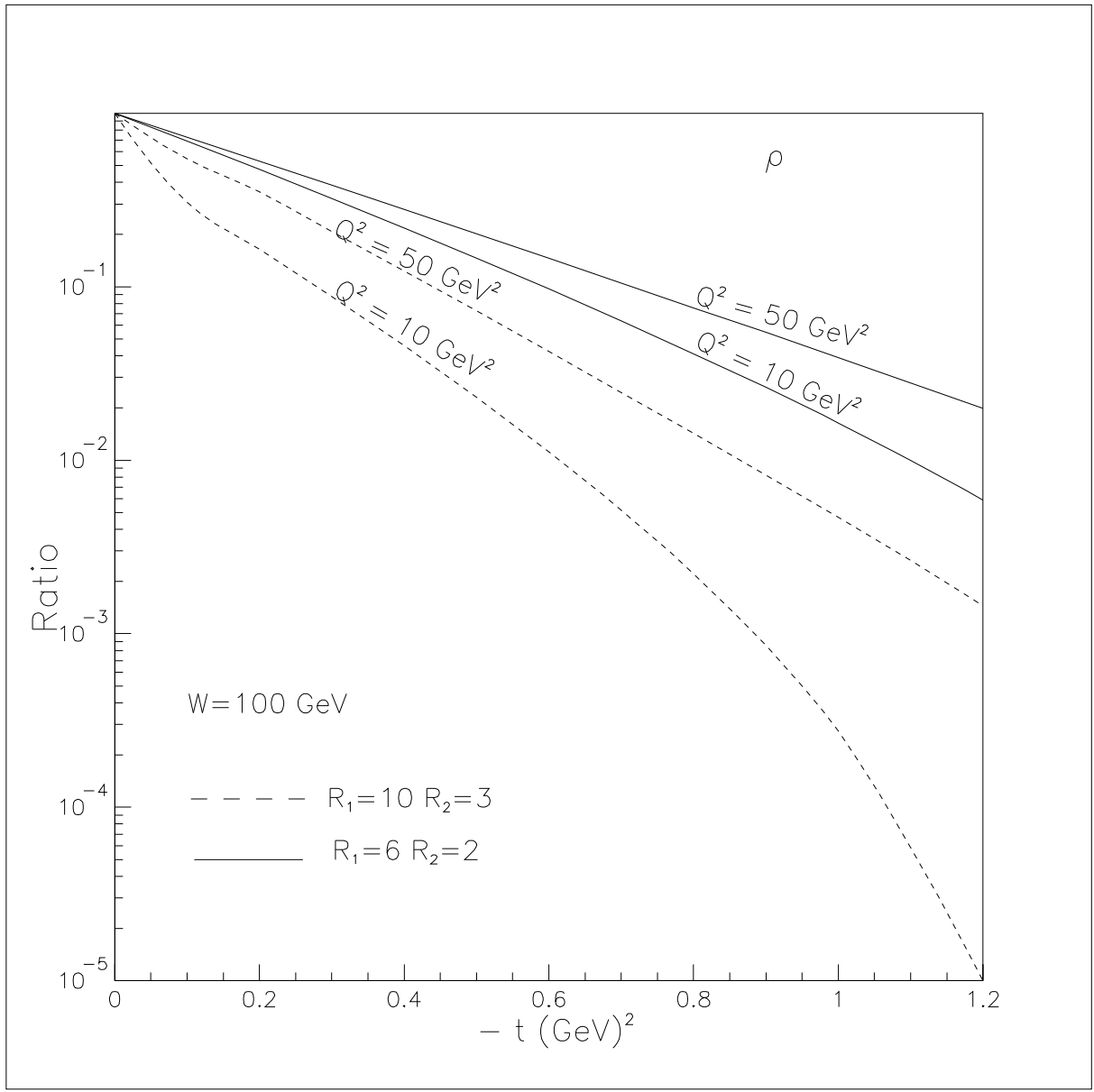,width=160mm}
\caption{$t$-dependence of the differential cross section for diffractive
$\rho$ production.}
\label{rhotslope}
\end{figure}

\begin{figure}
\epsfig{file=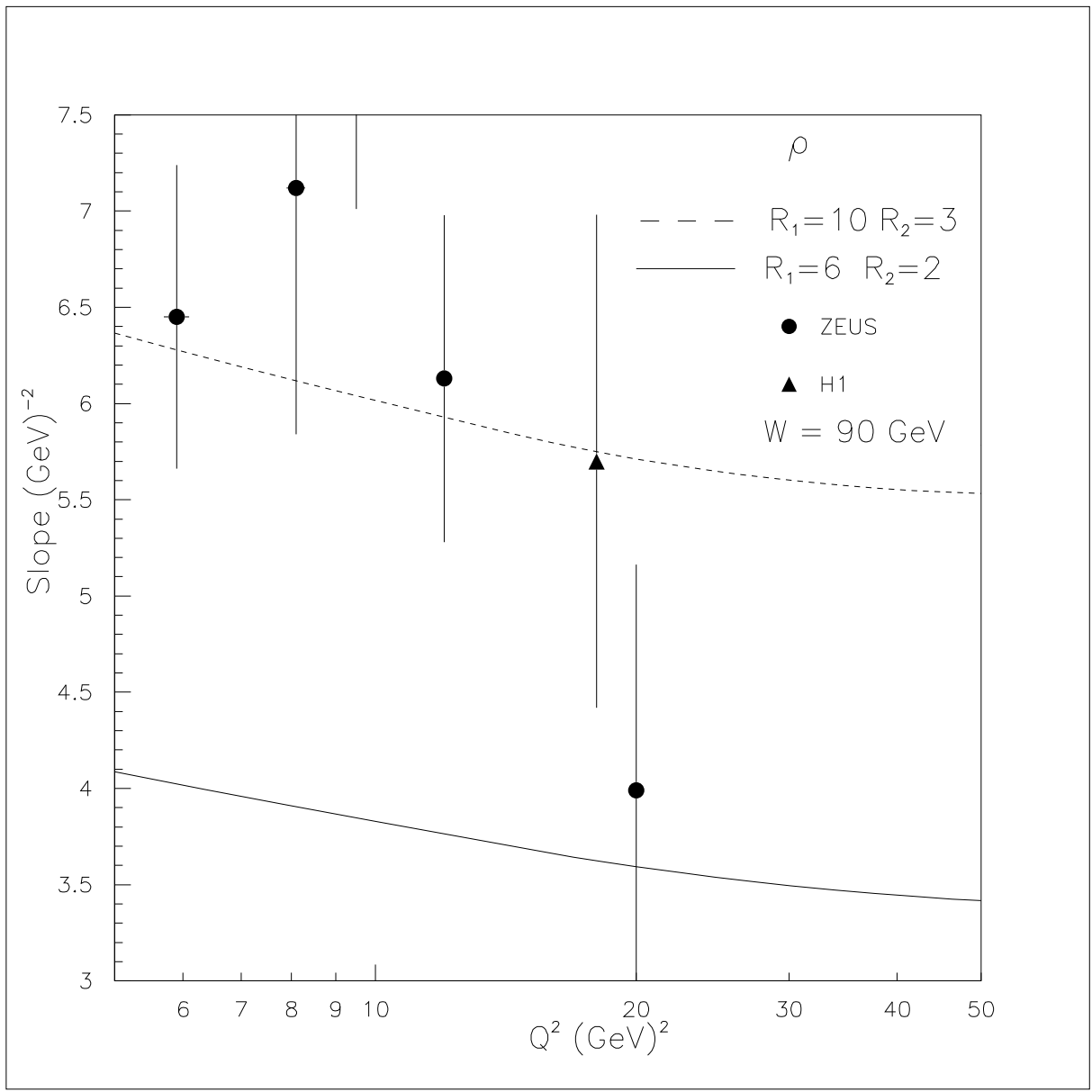,width=160mm}
\caption{$Q^2$ dependence of the slope at $t = 0$ for $\rho$ diffractive
production.}
\label{rhoqslope}
\end{figure}

\begin{figure}
\psfig{figure=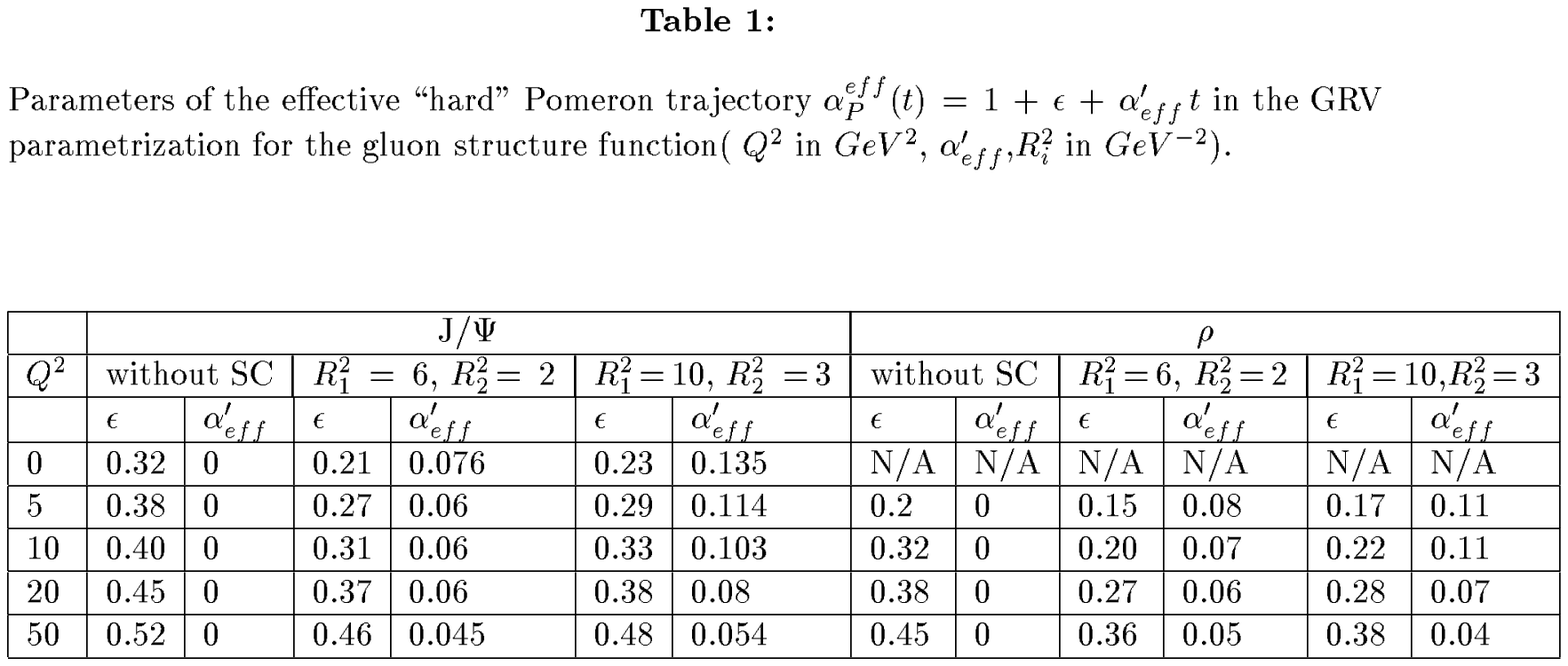,width=270mm,angle=90}
\end{figure}

\end{document}